\documentclass[10pt]{article}
\usepackage{geometry}                
\usepackage[parfill]{parskip}
\usepackage{amssymb}
\usepackage{graphicx}

\usepackage{amsmath}

\title{David Maurice Brink\\ 
{\small 20 July 1930 - 8 March 2021 \\
Elected FRS 1981} }

\author{C. V. Sukumar$^1$\footnote {candadi.sukumar@wadham.ox.ac.uk}, 
A. Bonaccorso$^2$ \footnote {bonac@df.unipi.it} \\{\small $^1$Wadham College, Oxford OX1 3PN, U.K}, {\small $^2$INFN, Sezione di Pisa, 56127 Pisa,  Italy.}}
\begin{document}
 
\maketitle

\begin{abstract}
David Brink was one of the leading theoretical nuclear physicists of his generation. He made major contributions to the study of all aspects of nuclear physics embracing nuclear structure, nuclear scattering, and nuclear instability. His wide ranging interests and interactions with theorists and experimentalists alike helped him in providing both theoretical analysis and interpretations and suggesting experiments. He had the gift of visualising complex problems in simple terms and provided clear analysis of the underlying processes. He was an expert on the use of semi-classical methods which provided an intuitively clear picture of complex phenomena. His research work and books are characterised by scientific clarity, transparency, and depth. David possessed outstanding skills in mathematical computation, and he was an expert on special functions, group theory, and the Feynman path integral method. David had many research students and collaborated with a large number of scientists from across the world, for whom he was a source of scientific and human inspiration and admiration. His most fundamental belief was that research was a means of trying to discover and understand the beauties of Nature and explain them in simple terms to others. His absolute belief in the value of truth and  his unselfish and generous attitude in sharing knowledge makes him an outstanding figure in contemporary Nuclear Physics. \end {abstract}

\begin{figure*}        
\begin{center}
\caption {David M. Brink}
\includegraphics[scale=.8]{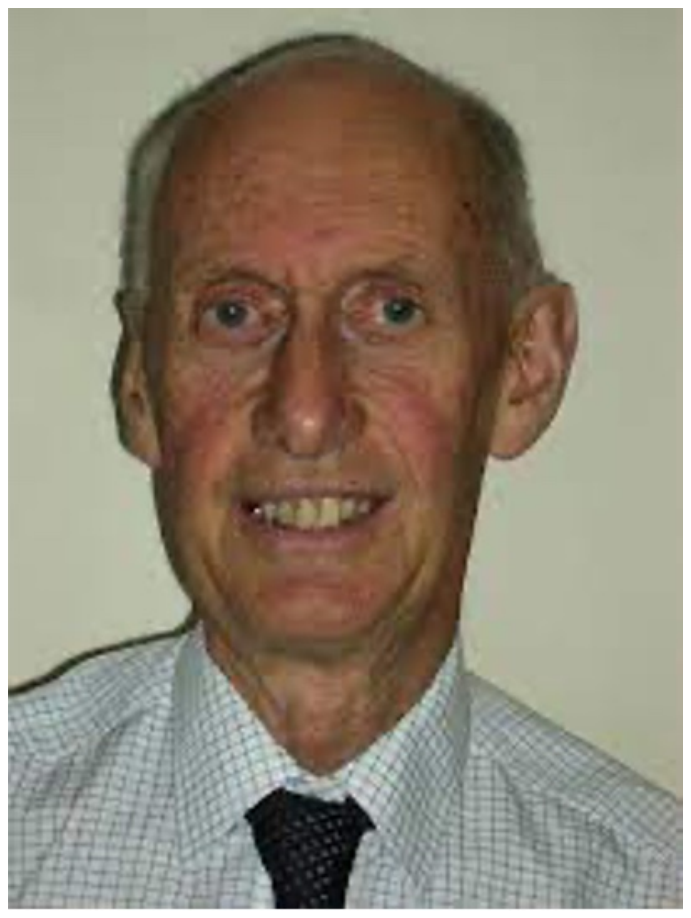}
\end{center}
\label{f1}
\end{figure*}          
                                                   
\section {   Early years  and family memories}

David Maurice Brink   was born on 20 July 1930 in Hobart, Tasmania. His father, Maurice Brink had been born in the village of Bjuv in Sweden in  1900. David's grandparents emigrated to Australia in July 1900. 
At the age of 14 David's father moved to Sydney where he trained to become an accountant. After this he went to Tasmania and joined an accountancy firm Wise, Lord  and Ferguson, where he eventually became  a partner. In 1929 he married Victoria Finlayson, (born in 1900). Her father David had emigrated with his parents from Scotland in 1884. They had an engineering firm in Devonport, Tasmania whose main activity was maintaining and repairing machinery for mining, shipping, and timber companies. David's grandfather and his colleagues built the first steam car in Tasmania and between 1900 and 1904 built nine vehicles including three passenger cars and one 12-passenger bus. David visited his grandparents often during vacations. He saw the casting floor and other parts of the factory and enjoyed playing amongst the remains of old steam traction engines. 

David was the eldest of three brothers. The Brink brothers went to a Quaker school in Hobart, Australia 1936 to 1948. David attended the University of Tasmania during 1948-51 studying Physics, Mathematics, and Chemistry, graduating with a BSc in December 1950 and was elected as a Rhodes Scholar at Magdalen College, Oxford, from October 1951. From February 1951 to September 1951 he studied for BSc Honours in Hobart but did not complete the course because he moved to Oxford in September 1951.
 
As a student at the University of Tasmania David joined the Hobart Walking Club. With this club he went on many trips to the interior of the island. When he arrived in Oxford he became a member of the Oxford University 
 Alpine Club. Its activities took him to the Alps where he climbed in the Valais and the Engadine in Switzerland. It was in Switzerland that he met his future wife Verena. Verena and David married in 1958 and had three children together. His love for walking was transmitted to his three children who continue to enjoy walking in urban, rural, and mountainous settings. While always very committed and absorbed with his Physics he was also a devoted husband and father, transmitting his joy for walking and travel to his family. He often helped his children with their homework and was very patient with them, even when they were not! Together David and his family travelled to, and lived in many countries across the world, where their horizons were broadened and they were introduced to the idea that there are many different ways of living and being. When his children had left home and travelled to other countries he would often be found in front of an atlas studying their exact whereabouts. 

David was very open minded and curious, always accepting other people's opinions and points of view. David and Verena were very close, shared everything and had full respect for each other. Verena was a wonderful host and the Brinks often organised tea and dinner parties for students, visitors, and their families. Verena also helped visitors find accommodation, and with other issues related to living in Oxford. They were also very generous in offering accommodation at their place whenever possible.

In Oxford David developed an interest in birds, initially just  birds he saw in Oxford, but when he travelled he always liked to look for birds and  made lists of species he saw. This curiosity in nature extended to other species as well, including trees. When in 1993 he moved to Trento, Italy, he became a member of the SOSAT, a branch of the alpine club, and went regularly with them on Sunday trekking trips.

 \begin{figure*}        
\begin{center}
\caption {David (right) in Tasmania 1950.}
\includegraphics[scale=.8]{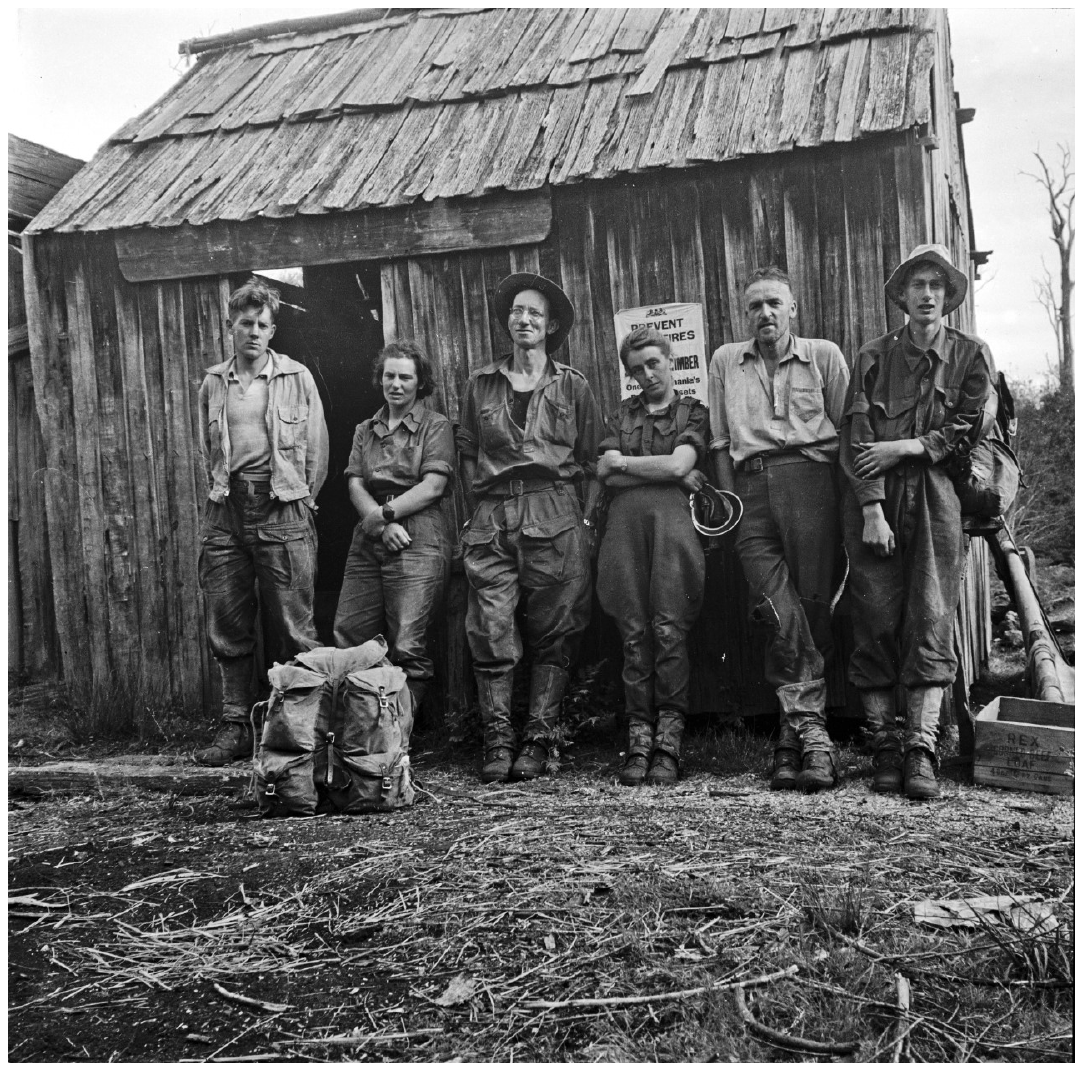}
\end{center}
\label{f2}
\end{figure*}          

\section{Graduate studies and Oxford beginnings}

David started his studies at Oxford in October 1951. When he arrived at Magdalen College there was no tutor in Theoretical Physics at the college. His maths tutor was David Kendal who sent him for tutorials to Jack De Wet at Balliol College. Jack asked David to read Von Neumann's book on the foundations of Quantum Mechanics in German. He also encouraged David to change his studies from a BA in Mathematics to a D. Phil in Theoretical Physics. Maurice H. L. Pryce (FRS 1951) was  the Wykeham Professor and head of the Theoretical Physics Department in Oxford from 1946 to 1954.  He was David's supervisor. Pryce was  also the part-time leader of the Theoretical Physics Division of the Atomic Energy Research Establishment (AERE) at Harwell, not far from Oxford, where nuclear theory was very much in the forefront and Rudolf E. Peierls (FRS 1945) was a consultant. At Harwell there was a very productive theory group  including Tony Skyrme  and J. P. (Phil) Elliott (FRS 1980). Skyrme organized regular informal meetings known as 'Skyrmishes'. Important papers in the latest journals were presented and discussed. Members of the group attended Oxford seminars and while the local group including
Roger Blin-Stoyle (FRS 1976), David Brink, and Pryce attended the Harwell meetings.  
Elliott gave some lectures at Oxford on Racah algebra. Later on his best-known work brought together the shell and collective models to explain rotational bands in deformed nuclei using the unitary group SU(3). During this time he wrote a long article in Handbuch der Physik with A. M.(Tony)Lane (FRS 1975) \cite{EL} on the shell model.

The foundations of David Brink's lifelong research, can all be found in his thesis "Some Aspects of the Interactions of Fields with Matter" \cite{thesis} which was submitted in May 1955. It is a remarkable document for its breadth and early contributions to the field of nuclear physics.  M. Pryce, his thesis adviser was interested largely in atomic spectroscopy but also studied the spectroscopy of nuclear energy levels. The advent of the shell model around 1950 opened the door to new theoretical approaches for understanding the properties of nuclei and applying quantum mechanical tools to calculate them. 
There was also a great interest in reactions involving heavy nuclei and which could only be treated by
statistical methods that had been developed much earlier. Brink's two-part thesis contained contributions to both areas, reflecting the interactions between the Harwell and Oxford groups. The first part  was inspired by the shell model and the second contains important contributions to the statistical theory of nuclear reactions. 

In the first part of his thesis, dealing with  Nuclear Structure, David analyzed the spectroscopic consequences of the nucleon-nucleon interaction acting on the valence nucleons in  nuclei close to the doubly-magic $^{208}$Pb. David  was able to estimate the order of magnitude of the interaction matrix elements from the properties of the deuteron. He also proposed treating the interaction through a density matrix expansion. This would figure prominently in later work in the field.

The second part of his thesis dealt with reactions involving heavy nuclei. It was probably inspired by the work of  experimental group at Harwell. There, a Van de Graaff accelerator was used to measure  energy levels, moments and transition rates in nuclei. David was also fortunate to have contact with the strong experimental group working on neutron resonances. While  David  was working on gamma widths of neutron resonances he benefited from contacts with Prof. Hughes \cite{hh} and Prof. Weisskopf who were visiting Oxford. Weisskopf was very much interested in applying the detailed balance theory to nuclear reaction and interactions with him must have influenced David because at the end of the thesis he acknowledges discussions with Victor Weisskopf. The first subject in this part was the theory of inelastic scattering on deformed nuclei. David constructed a theory for the excitation of rotational
bands in deformed nuclei based on two new ideas, namely Bohr's model of deformed nuclei and the optical model of Weisskopf et al. \cite{optmod} published the previous year. David was able to carry out the calculations to a point where the relative importance of this mechanism in the total cross section could be estimated. This was an impressive achievement at a time before computers were available to carry out the full calculations.

The final section of his thesis deals with the decays of the compound-nucleus resonances produced in reactions on heavy nuclei. The formulas he presented here are still in use for modeling the spectra and reactions in heavy nuclei \cite{Capote}. The best known is the formula for gamma decay rates in compound-nucleus resonances. This formula is based on a treatment widely known as the "Brink-Axel" hypothesis. At a fundamental level, the theory was derived from the principle of detailed balance which Weisskopf had used very successfully in other contexts. The principle gives a formula to relate decay
rates to absorption cross sections in the inverse reaction. The Brink-Axel hypothesis simply states that the absorption cross sections for gamma radiation on excited states of heavy nuclei can be estimated by the corresponding cross sections on the ground states. Axel and Brink worked independently. Peter Axel's paper appeared in 1962 \cite{PA}. The important statement is made on page 101 of David's thesis and is expressed in equation (11) of Axel's paper. The prediction of the statistics of the widths of nuclear resonances, based on the generalization of the central limit theorem which David had learned about in his statistics course in Tasmania. David published the results in \cite{dmb1} where he showed the close connection between the shell-model description of the giant dipole resonance and the collective model of Goldhaber and Teller \cite{GT} and Steinwedel and Jensen \cite{SJ}. After his paper, theory of the giant resonances used the shell model as a starting point. Confirmation of the Brink-Axel hypothesis first came from the  Berkeley  experiments in 1981 \cite{New}.

The last part of thesis has formulas related to another important topic in compound-nucleus theory, the fluctuations in decay widths of individual resonances. Here, David speculated that the fluctuations would follow a chi-squared distribution with one or two degrees of freedom. This is borne out experimentally and is now considered one of the hallmark properties of the compound nucleus.  It also became a part of
random matrix theory in mathematical physics. Unlike the early parts of the thesis, David never published the parts on compound-nucleus decay widths.  However, physicists at the Harwell Laboratory knew about David's results and J.E. Lynn explained them in his book \cite {Lynn}. Unfortunately, David's treatment of fluctuations was not recognized until very recently \cite{Geo} and the distributions are known today under other author's names \cite{PT}. 

 \begin{figure*}        
\begin{center}
\caption {David and his children (left to right), Barbara, Thomas, and Anne-Katherine 1969.}
\includegraphics[scale=.4]{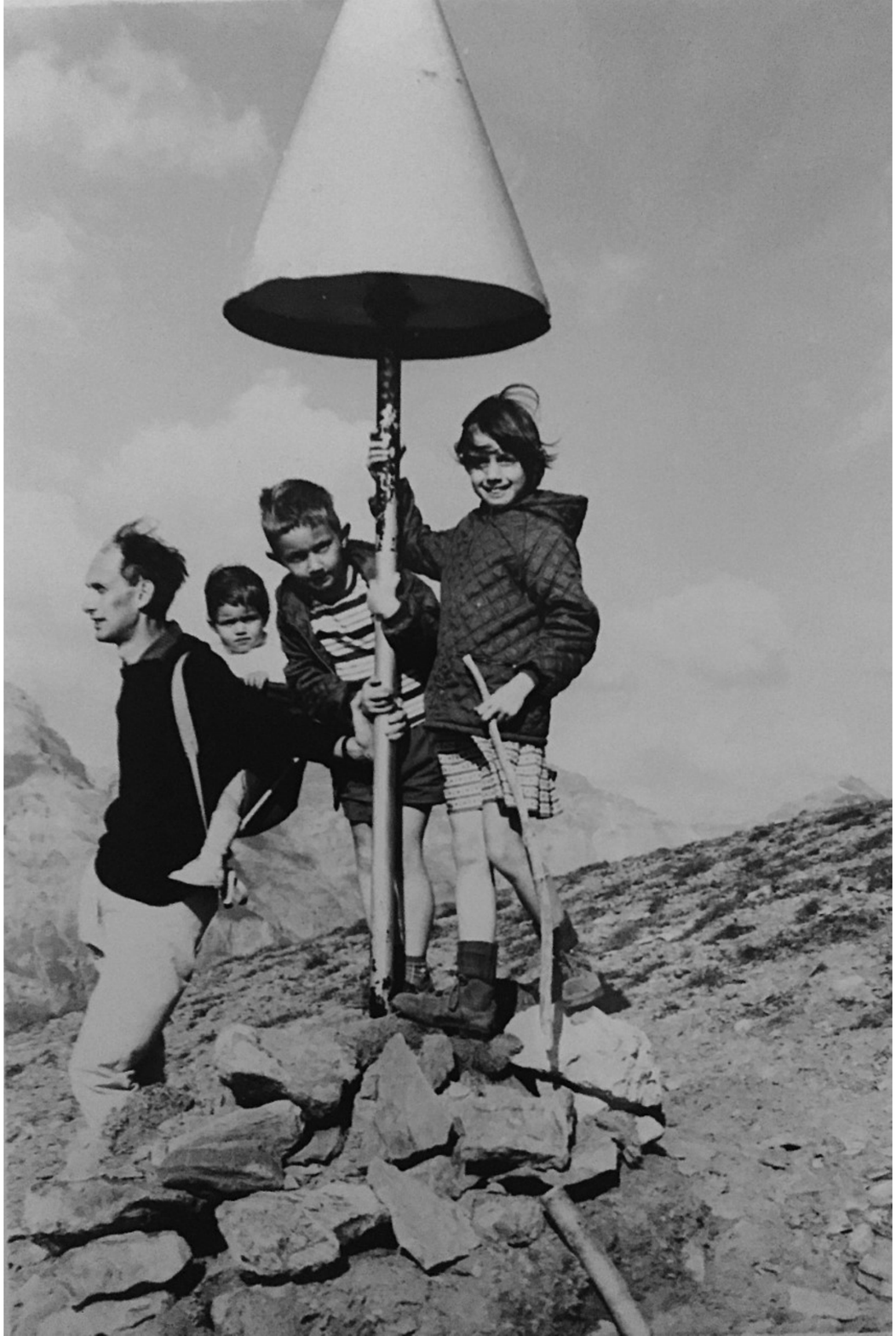}
\end{center}
\label{f3}
\end{figure*}

\section{Research areas}

David's interactions with the physicists mentioned earlier were reflected not only in David's thesis but also in his early publications.  One paper \cite{Rose}, which dealt with angular momentum couplings and angular distributions of $\gamma$-rays and other particles, is still the "Bible" most experimentalist use when they analyse their data, as we have been told by Peter Butler (FRS 2019) (Liverpool) and Yorick Blumenfeld (Orsay), and others. Early in his research career David wrote the textbook {\it Angular Momentum} \cite{b1} with Ray Satchler. This textbook was prominent among several texts published in this time period.  It was widely used by graduate students and post-graduates working in nuclear theory. David also published a book on {\it Nuclear Forces} \cite{b2}. 
      
\subsection{Effective interactions and calculations tools}

In his thesis David had laid the basis for the use of effective interactions in the calculations of matrix elements for nuclear structure studies. The idea was greatly advanced  in three later papers. The first proposes a gaussian form for the effective nucleon-nucleon interaction known as the "Brink-Boeker"   interaction \cite{BBO} that all nuclear physicists have used at least once in their lives. This paper was very influential at the time and was later developed by Gogny and collaborators in the interaction that is widely used even today \cite{Gogny1,Gogny2}.

In 1959 Tony Skyrme proposed  modelling the effective interaction between nucleons in nuclei by  a short-range potential, an idea which is useful in nuclear structure and the equation of state of neutron stars \cite{sky}. The Skyrme force is an effective interaction depending on a small number of parameters whose strength could be fitted to reproduce various bulk properties of nuclei as well as selected properties of some nuclei, especially the doubly magic nuclei. At the beginning of the 1970s David was a frequent visitor to the Theoretical Division at the Institut de Physique Nucl{\'e}aire, Orsay where his sixty-fifth birthday was celebrated (figure 4). The work done there produced two papers with Dominique Vautherin \cite{vauth1,vauth2} which were the basis for the intense use of the so-called Skyrme interactions, in all their many present variants. The papers revived a general interest in using Skyrme's parametrization of the nucleon-nucleon interaction to calculate nuclear binding energies, and later to other aspects of nuclear structure. In effect, the interaction is treated as an energy-density functional theory in the spirit of the Kohn-Sham theory in condensed matter physics.

The Hartree-Fock calculations in \cite{vauth1} for spherical nuclei used Skyrme's density dependent effective interaction. This seminal paper showed how the Skyrme force could be used to make accurate calculations of certain nuclear properties and Vautherin and Brink developed these ideas further in a series of papers which had a strong impact on nuclear structure calculations. T. Otsuka comments: ``The paper \cite{vauth1} has had a huge impact, as verified by the number of citations $>$2000. In nuclear theory, papers having the citation index $> $1000 are rather few, which implies how important the Vautherin-Brink paper is. This year is the 50 year anniversary of this paper, and it is amazing that the basic formulation within the mean-field approach has not changed too much, implying that the scheme presented in this paper is so solid".
 
The calculations of Vautherin and Brink were extended by many other physicists during the subsequent period. In particular at Oxford, Micky Engel, Klaus Goeke and Steve Krieger, together with Dominique Vautherin derived the energy density using a Slater determinant where the single particle states were no longer invariant under time reversal, as it is in the Hartree-Fock method. With the Skyrme interaction the TDHF approach leads to an equation of continuity for the single particle density \cite{vauth2}. This paper showed how Dirac's time-dependent Hartree-Fock theory could be applied to nuclear dynamics in a light nucleus. In the year immediately following the publication, the theory was applied to collisions involving a large number of nucleons \cite{bonche}, showing that the method would be a powerful one for heavy nuclei as well. The method is justified as a time-dependent density-functional theory, and it remains in widespread use.
 
In 1973 Ica Stancu came to Oxford  as a post doctoral fellow and worked with David on heavy ion reactions in deriving the interaction potential of two $^{16}$O nuclei starting from the Skyrme energy density formalism \cite{SB}. They included the previously ignored tensor part of the Skyrme interaction. Along with an additional effort from Hubert Flocard at Orsay, the Skyrme HF calculations yielded single particle levels of spherical closed nuclei \cite{SB1}. The role of the tensor force is to contribute to the spin-orbit splitting of the single-particle levels. For spherical closed shell nuclei the effect turned out to be small. Later it was found that in spherical spin unsaturated nuclei it makes a dramatic
difference, giving the correct order of single particle levels, as, for example, in the Sn isotopes \cite{SB2}. Many experiments on neutron-rich nuclei since 2006 have shown that the Skyrme formalism including 
the tensor force was the simplest way to describe the shell evolution of neutron-rich or proton-rich
nuclei and indicated new magic numbers.   

\subsection{Heavy-ions and Semi-classical methods in Nuclear Physics}

As tandem accelerators and cyclotrons were built to study heavy-ion Physics, David started an intense collaboration with the experimentalists at the Department of Nuclear Physics in Oxford. The accelerators were used to study heavy-ion elastic scattering and direct reactions such as transfer and measure masses and perform spectroscopy of neutron-rich matter. In those years  semiclassical methods were widely used in the Nuclear Physics community to analyse  data. They were particularly appropriate for heavy ions because of the high incident energies and the large impact parameters involved. Thus David started the Oxford school on the subject, more or less parallel in time to the Copenhagen school of Broglia and Winter and collaborators. At that time, these heavy-ion reactions were analyzed through the partial wave expansions of the colliding partners, a methodology that was computationally demanding and giving little insight to the underlying dynamics. David's  semi-classical treatment of the collision was much simpler. Some of  the early papers on the theory of peripheral reactions  were based on his student's thesis,  including Hashima Hasan and Luigi Lo Monaco \cite{HB,lmb}.  

David's investigation of the kinematical effects in such reactions, for which there was concrete experimental evidence from the work of Peter Twin (FRS 1993) and his collaborators at Liverpool,  became a key element for experimentalists.  In the paper by the title ''Kinematical effects in heavy-ion reactions'' \cite{kin} David   introduced a "semi-classical amplitude" \cite{b3} that could be used in DWBA-like calculations of transfer \cite{t1} and proposed a matching condition to predict a large reaction cross-sections, a condition that was beautifully adapted to understand spin-polarization experiments. He showed that energy and angular momentum couplings  in heavy-ion reactions led to very selective matching rules by which high angular momentum single-particle states could be populated. High angular momentum single-particle states  sometimes appear as low-lying continuum resonances. They have been studied by the method of transfer-to-the-continuum \cite{bb} which has helped disentangle single-particle from collective degrees of freedom and has also been applied in the so called "surrogate reactions" as a substitute for  free neutron beams.

Semi-classical ideas have been  helpful in studying  breakup and dissociation of weakly bound radioactive ions including  halo nuclei and other such unstable nuclei  whose dynamics is rather involved and difficult to study experimentally due to the very low intensity of beams. David, Angela Bonaccorso and her students got heavily involved in this new physics from the '90s on, with a long series of papers (see \cite{bb1} and references therein), conference contributions, meeting organization, some of them at the ECT* in Trento, spanning the last forty years of David's career. Finally it has recently been shown \cite{jinme} that the semi-classical treatment of breakup by David and his collaborators is fully consistent with a  quantum mechanical treatment.

David studied microscopic models for the real and imaginary parts of the ion-ion optical potential to be used in elastic scattering calculations with Ica Stancu. He also studied fusion with Neil Rowley and N. Takigawa. David and Takigawa developed a semi-classical reaction theory with three classical turning points which explained the anomalous large angle scattering of ${\alpha}$ particles as a quantum-mechanical interference between the barrier wave and the internal wave, thereby providing an intuitively clear picture of a complex phenomenon underlying nuclear reactions in terms of classical and quantum ideas. David, Vautherin, and M.C. Nemes studied the effect of intrinsic degrees of freedom on the quantum tunnelling of a collective variable. This work was further developed by other theorists including Kouichi Hagino who studied the deviation from adiabaticity in quantum tunnelling with many degrees of freedom.

David met Uzi Smilanski in Munich when they were both there on sabbatical. Both had worked on semi-classical approximations and gave a joint series of lectures on this topic. David was concerned that the standard WKB method was insufficient to explain tunnelling through a barrier and was particularly bad near the barrier top. David and Uzi applied the uniform semi-classical method evolved by Michael Berry  (FRS 1982) to successfully address the problem \cite{berry}. Uzi remembers David as a physicist with excellent intuition and an ability to grasp the essence of a problem before cracking the problem with rigorous mathematics and complex computation. 

David, Massimo di Toro, and Alberto Dellafiore developed a semi-classical description of collective responses with a mean field approach paving the way for a study of the dynamics of a nuclear Hartree-Fock fluid. When the national heavy-ion laboratory started in Catania (LNS-INFN) around an advanced superconducting cyclotron, David was a reference point for simple physics suggestions. 

\subsection{Path integral methods in Nuclear Physics}

David's expertise with semi-classical methods for tackling quantum problems naturally led him towards the Feynman path integral approach to quantum mechanics which was based on a Lagrangian approach. Hans Weidenm{\"u}ller had met David at various conferences in the 1950s and 1960s and spent 1977-78 on a sabbatical in Oxford. During this period David and Hans worked on the application of the Feynman path integral method to the study of the heavy-ion reactions and developed the Influence Functional approach to this problem which David and his collaborators later used to establish master equations. Hans remembers that at a summer school a few years later David delivered a series of lectures on nuclear reactions. In the first lecture he developed the topic using a dozen transparencies and in subsequent lectures used the same transparencies in a different order to display and illuminate aspects of the topic that had gone unnoticed before. Hans remembers it as a display of the combination of simplicity and depth that were hallmarks of David's approach to Physics.

The path integral method was particularly well suited for studying problems with many degrees of freedom in which classical description in terms of trajectories was good for some degrees of freedom but not for all. Coulomb excitation in heavy-ion collisions is an example where the relative motion of the ions could be described in terms of coulomb trajectories but the excitation of the quantum states of the ions had to be treated using quantum mechanics. David and Sukumar \cite{sd1} used the Feynman path integral method to evolve a systematic way of arranging the correction terms for the quantum amplitudes for processes involving coupled degrees of freedom where the description in terms of classical trajectories was good for some degrees of freedom. David, Sukumar, and Fernando Dos Aidos used this method to provide corrections to the primitive semi-classical amplitude for Coulomb excitation of heavy-ions. Sukumar and David used the path integral method to describe spin-orbit coupling effects and together with Ron Johnson at Surrey and his group successfully explained the experimental data on polarization effects.

\begin{figure*}        
\begin{center}
\caption {David and his wife Verena, May 2018.}
\includegraphics[scale=.4]{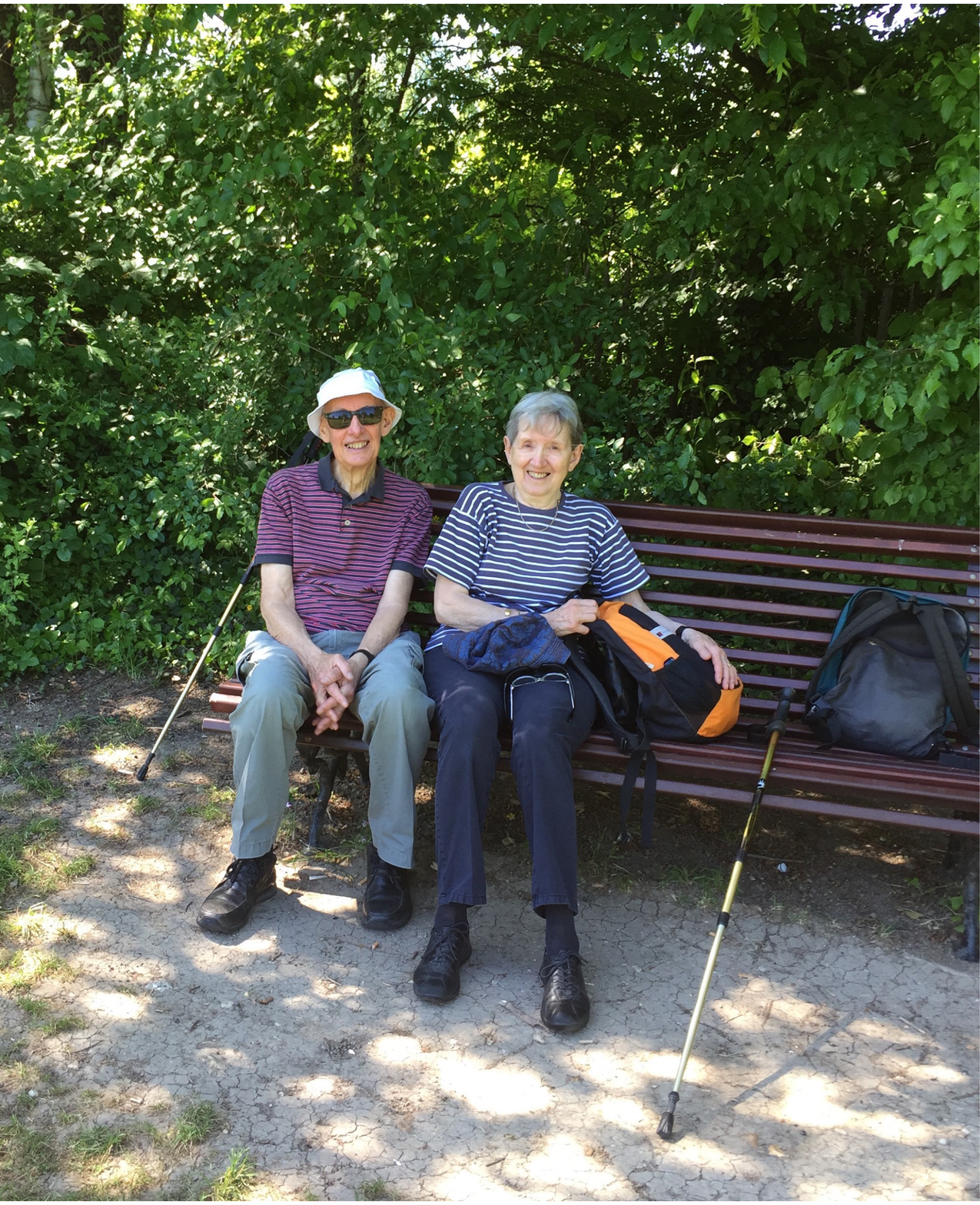}
\end{center}
\label{f4}
\end{figure*}          

\section{Other topics}
David was very quick at grasping the core of a Physics problem and putting it in simple, calculable terms. Often the problem required somewhat involved analytical calculations, but he was a master of that. Thus anytime a visitor went to Oxford with a new problem, David would start a very successful line of research which he often followed up with his graduate students.
  
\subsection{Cluster models}
It happened for example with the cluster  model physics, starting with the seminal paper \cite{BW}. This paper developed the generator coordinate method of Hill and Wheeler  \cite{hw} to produce a practical tool to reduce
the many-particle Hamiltonian to an ordinary Schr\"odinger equation for a collective variable. Thus the nuclear cluster model was related to the shell model. To treat nuclear states in such different circumstances, a formulation which includes clustering at one extreme and shell structure at the other extreme was needed. David proposed microscopic multi-${\alpha}$-clusters treating four nucleons with different spin-isospin states as a single particle orbit. Under anti-symmetrisation of nucleons the cluster model wave-functions approximate shell model functions and enabled the description of both cluster and shell model structures in a unified way. Their approach was adopted and is in widespread use even in present-day nuclear theory. The main applications up to now are on spectroscopy and large-amplitude collective motion. 

Y. Suzuki's work on the cluster model was largely inspired by David's paper on "Do alpha clusters exist in nuclei?" \cite{br3} presented at a meeting in Tokyo in 1975. This paper contained all the essential components needed in the alpha particle model, the microscopic theory beyond the shell model description based on many-particle many-hole excitations, the relation between the resonating group method GCM, the equilibrium arrangement of clusters, extension of the Hill-Wheeler method, the angular momentum projection, and the Slater determinant technique for evaluating matrix elements. Suzuki remembers that David never forgot to mention that the original model was proposed by H. Margenau and C.Bloch \cite{mar, br2,br1}.

At the Varenna School in 1955 David met S. Yoshida from Japan and they discussed  inelastic scattering of protons and neutrons by deformed nuclei. By chance David had a chapter in his thesis on this topic and Yoshida had been studying the same subject. This interaction with Yoshida helped David to develop strong connections with nuclear theory groups in Japan over many years.

\subsection{Bose-Einstein condensation of atoms}

During his period as Deputy Director of ECT* in Trento, 1993-1998, David interacted with many members of the Physics Department in Trento. One such interaction with Sandro Stringari led to David's interest in Bose-Einstein condensation of alkali atoms in magnetic traps \cite{Sandro}. Sukumar and David \cite{sd2} developed an approximate method for calculating the rate of escape from the magnetic trap thereby enabling an estimation of the duration for which the condensate atoms can be held in the trap as a function of the ultra-cold temperature and the strength of the magnetic field.
 
\subsection{Miscellaneous }

David was interested in the role of pairing interaction in finite nuclei and this led to the study of nuclear superfluids. His book with R. Broglia \cite{b4} is considered to be a wonderful exposition of this subject. David's knack for  explaining detailed Physics in a simple and clear manner is abundantly evident in this book. In the 1990s Ica Stancu raised David's interest in the quark structure of exotic hadrons named tetraquarks, a system of two quarks and two antiquarks, and studied the stability of such systems containing heavy quarks/antiquarks in a QCD inspired quark model. Even though David had not worked on the Interacting Boson Model (IBM) he nevertheless provided supervision for doctoral students such as Martin Zirnbauer who chose topics in this field. He also supervised Hans Peter Pavel's thesis on Schwinger pair production in a flux tube model containing a chromomagnetic field.

\section{Teaching and administrative roles}

David's doctoral students remember him for the gentle way he corrected them when they had made errors. Many of the students learned from him how to take a critical approach to their results and how it is possible to look at a complex problem from several different viewpoints and find the one that gives the best physical insight. They also remember the immense support he gave to their research and pastoral care. Many graduate students also remember how much they had learned from the courses he taught at Oxford and at Summer schools. His book  with Satchler  \cite {b1} and paper  with  Rose \cite{Rose} on angular momentum algebra  were found to be of immense value in formulating and tackling problems in Nuclear Physics. Many researchers and students who met David were astonished that someone with such towering achievements could be so humble, nice and honest.
David was very open-minded and we report a number of episodes to illustrate this aspect of his character.

Future Nobel laureate Prof. Tony Leggett remembers:
" My undergraduate major at Balliol was in Greats (classical languages, ancient history and philosophy) and I was set to graduate (and eventually did so) in the summer of 1959. Towards the end of the academic year 1957-1958, partly encouraged by the post-Sputnik cultural swing towards science in the UK, I conceived the ambition of taking a second undergraduate degree in physics and perhaps eventually making my career in academia in that subject. Given that I had essentially no meaningful exposure to physics at the high-school level and only a  brief and  informal exposure to any kind of mathematics beyond simple differential calculus (I'm not sure that I had even had that), such a drastic change of academic direction was extremely unusual, indeed at the time almost unheard-of. My first concern was to find a higher education institution which would accept me for it  and I rapidly concluded that my only hope was to apply to my existing Oxford college, Balliol. David had just recently become the college's first tutor in theoretical physics (most Oxford colleges did not have such a thing in 1958), so it fell on him to take the decision on my application. To this end he asked me to read over the summer vacation a few chapters from the book
"What is Mathematics?" by Courant and Robbins \cite{CR}, perhaps the most beautiful presentation I have ever seen of mathematical topics for the  layperson. When I returned to Oxford in the Fall of 1958 he gave me an informal mini-exam on that material, and on the basis of my performance decided to recommend to Balliol to accept me. In the event I did my physics degree at Merton, who offered me a scholarship, but since they did not at the time have a tutor in theoretical physics David played that role for me for much of the two years which it took me to complete the degree. I think it is virtually certain that had he made the opposite decision, I would never have had a career in physics, and I am profoundly grateful to him for the imagination he showed in going beyond my formal academic qualifications."

Another story comes from Paul Stevenson:
"I was called up for interview at Balliol in December 1991. The office I was in for that interview was David Brink's office, above the Senior Common Room.  In the interview were me, David Brink, David Wark, Jonathan Hodby (those three there for physics) and Bill Newton-Smith (for philosophy). 
  I don't remember all the questions. 
I do remember that David Brink showed me a postcard and asked me what, physically, was wrong with the picture.  It was a Japanese style print with a mountain in the background and a lake in the foreground.  There was a reflection of the mountain in the lake, but it was off to one side.  I saw what was wrong, and struggled to articulate it in the language of a physicist, and in the end David prompted me by asking what is particular about an incident light ray, a reflected light ray, and the normal to the surface at which it is reflected and I said the right thing - that they are all in the same plane. I was duly accepted to Balliol and spent three years there studying physics". 
Danny Chapman remembers:
 "I don't think I'll ever forget the "sense"  of David Brink's tutorials, and of being in the presence of such a sharp and insightful mind. I remember being quite inspired once when my fellow student had tried to answer a question in what I thought was an odd and probably wrong way, ending up with a sum, which he then attempted to turn into an integral, which didn't work out. Rather than saying "don't do it like that, do it like this", David was able to continue from there and make it work, which was a really positive experience and encouragement to follow every path to its end.  I feel lucky to have been at Balliol when he was there."

Angela Bonaccorso remembers daily life as one of David's students: At the Department of Theoretical Physics there was a coffee room where coffee was served between 11:00 and 11:30. We would try to be there on time to sit around David who would be chatting with other senior members of the department or some visitor. There would always be someone bringing up some interesting and challenging new problem. Everyone gave an opinion, the atmosphere was competitive. Most of the time David would win the argument and his students felt very proud.Not all supervisors were so nice, helpful, and respectful of us as David was. But it was not at all easy to be David's student. First of all we needed to have detective skills. David was very busy and very elusive. In those days there was no email or SMS. The only way to be sure that he was inside was to look for his bicycle. If the bicycle was outside we would knock at the door of his office and if we were lucky he would answer and let us in. In spite of all his many commitments we always managed to have at least one chat per week with him. Another reason why it was not easy to be his student was that David had a very original way of understanding things and finding the way out of problems. During our conversations often he would stop talking and be silent for five to ten minutes, rubbing his hand on his forehead. Then he came up with some equation, or a drawing or something like that and he would tell us: I think it is like this...I think we should get something like that...etc. I (we) would stare at him speechless and in wonder. Where did the 'oracle' come from? Most of the time this was the end of the meeting. I (we) left his office rather puzzled, worked desperately hard for one week and if we had managed to understand his line of thought, after pages and pages of calculations, we would find exactly what he had predicted. We all knew it was like that, we all passed this information on to each other, generation after generation: listen to David, he is always right, just try to reproduce the miracle of his craftsmanship in physics.

A further proof of how much busy David was and how precious was for everyone the time spent in conversation with him can be found in the comment Gerry Brown made in his review  for Science  \cite{gB} of the Proceedings of the Varenna summer school \cite{br2} :
'Let me draw special attention also
to the article of David  Brink, "The alpha-particle model of light nuclei," which is one of the most beautiful developments in this subject. Brink likes to sit on his work for years and, on the whole, doesn't even answer letters inquiring about it, so that one must either adopt the expedient of traveling to Oxford to talk with him, or invite him to lecture at summer schools. Both are worth while.'

David was a pillar of Balliol college and Department of Theoretical Physics for decades, an immensely popular tutor and supervisor, a cheerful and always helpful colleague, and a wonderful guide to younger colleagues and administrative staff who happened to be working with him. David had another long and distinguished career in Italy after he left Oxford. Following an invitation from Renzo Leonardi he moved to Trento as full professor of History of Physics and helped in establishing the ECT*, European Center for Theoretical Studies in Nuclear Physics and Related Areas. The Nobel laureate Ben Mottelson was the founding director and David the vice-director, while Renzo Leonardi was the Scientific Secretary. In the five years David spent at Trento he took care of organising various technical aspects of the secretarial offices, library, computer center and visitor hospitality. At the same time he gave very productive contributions to workshops with his constant presence, his huge knowledge of nuclear physics and stimulating discussions. The superb reputation and international standing of this extremely important European initiative is undoubtedly due in large part to David's wisdom in its crucial, formative years.

\begin{figure*}        
\begin{center}
\caption{David's sixty-fifth birthday celebration. Orsay, 1995.}
\includegraphics[scale=.5,angle=90]{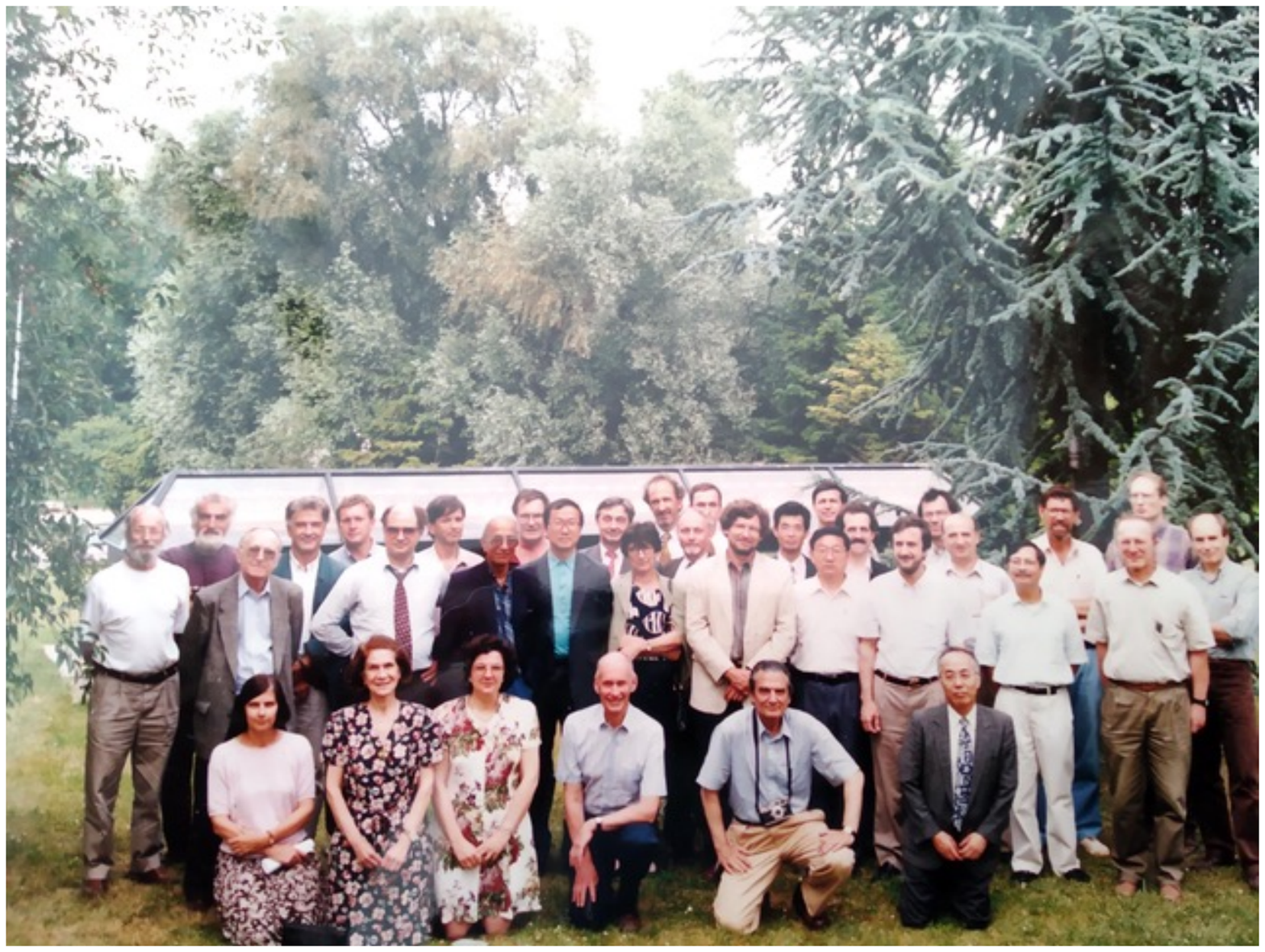}
\end{center}
\label{f5}
\end{figure*}
          
\section{Career, Honours and Awards}

1954-55 Royal Society Rutherford Scholarship.\\
1957-1958 Instructor at the Massachusetts Institute of Technology (MIT).\\
1958 Fellow of Balliol College and Lecturer in Theoretical Physics, Oxford. \\
1976-1978 Vice-Master of Balliol College.\\
1981 Fellow of the Royal Society.\\
1982 Rutherford Medal of the Institute of Physics.\\
1988 H. J. G. Mosley Reader at Oxford. \\
1990-1993 Senior Tutor, Balliol College, academic planning and  administration, Oxford.   \\
1992 Foreign member of the Royal Society of Sciences, Uppsala.\\
1993-1998 ECT*, Trento, Vice-Director .\\
1993-1998 Full professor of History of Physics, University of Trento.\\
2006 Varenna Conference on Nuclear Reactions  dedicated to him.\\
2006 Lise Meitner prize of the European Physical Society shared with H. J. Kluge.

\vskip 30pt

Visiting scientist at : 
\begin{itemize}
\item 
Niels Bohr Institute 1964, 
\item 
University of British Columbia 1975, 
\item 
 Institut de Physique Nucl{\'e}aire d'Orsay 1969 and 1981-1982, 
\item 
The Technical University of Munich 1982, 
\item 
University of Trento 1988, 
\item 
University of Catania 1988, 
\item 
Michigan State University 1988-1989.
\end {itemize}

\section{Acknowledgements}
The authors are greatly indebted to the Brink family for sharing with them  private memories and photographs and for a critical reading of the manuscript. A  large number of  friends and colleagues, too many to be individually mentioned, contributed with their appreciation of David's life and scientific career. Ica Stancu and Sharon McGrayne Bertsch read and commented the manuscript. One of us (AB) gratefully acknowledges George F. Bertsch for his help in digging out from David's thesis and early work the roots of several founding pillars of modern Nuclear Physics.

\end{document}